\documentclass[12pt]{iopart}

\pdfoutput=1

\usepackage{graphicx}
\usepackage{color}
\usepackage{amssymb}
\usepackage{bbm}

\newcommand{\ket}[1]{\ensuremath{\left|#1\right>}}

\newcommand{\figwidth}{3.2in}
\newcommand{\largefigwidth}{6.4in}

\begin{document}

\title[Quantum Chessboards in the Deuterium Molecular Ion]
{Quantum Chessboards in the Deuterium Molecular Ion}

\author{C R Calvert, T Birkeland$^*$, R B King, I D
Williams, J F McCann - \emph{{\today}}}
\address{School of Mathematics and Physics, Queen's University
Belfast, BT7 1NN, UK}%
\address{$^*$Permanent Address: Department of Mathematics, University of Bergen,
N-5007, Bergen, Norway}

\ead{tore.birkeland@math.uib.no}

\begin{abstract}
We present  
a new algorithm  for vibrational control in deuterium  molecules 
 that is  feasible with current experimental technology. A pump mechanism is used to create a coherent superposition of the D$_2^+$ vibrations. A short, intense 
 infrared  control pulse is applied after a chosen delay time to create selective interferences. A `chessboard' pattern of states can be realized in which a set of even- or odd-numbered vibrational states can be selectively annihilated or enhanced. A technique is proposed for experimental realization and observation of this effect using 5 fs pulses of
$\lambda=790$ nm radiation, with intermediate intensity ($5 \times 10^{13}$ W
cm$^{-2}$).
\end{abstract}


\section{\label{sec:level1}Introduction}

The ability to effect quantum control at the molecular level
has many potential applications, such as  control of
reaction dynamics \cite{zewail}, selective electron localization
\cite{Vrakking_Nature_dtwoplus_loc}, and  applications
in quantum information processing
\cite{Domhnallvibcontrol}. In order to characterize control
mechanisms and afford useful comparison between experiment and
theory the hydrogen diatomic (and its ion, H$_2^+$) is
often used as a test-bed. Such fundamental systems are extremely useful for identifying basic physical mechanisms because of their simple electronic structure. The heavier isotopes of hydrogen, such as deuterium and deuterium hydride are equally
accessible for quantum simulations and execute
slower nuclear motion, thus deuterium (D$_2$) has become the
preferred target for time-resolved studies.

In recent years, ultrashort laser pulses have been used to
initiate, image \cite{Trump,Ergler} and control \cite{Dissociation
Control,Niikuraexpcontrol} the ultrafast dissociation of D$_2^+$,
with the localization of the remaining bound
electron also  manipulated \cite{Vrakking_Nature_dtwoplus_loc}.
However, such control of fundamental molecular motion has not
been limited to dissociation dynamics and progress has been made
on studies of bound wavepacket motion, with control mediated by
interactions occurring on timescales shorter than the motion
itself. For example, coherent rotational motion in molecules can
be induced and controlled \cite{Lee_Control,Fleischer} in diatomic systems using intense-field pulses on sub-rotational timescales, with the mapping of rapidly rotating (hydrogenic) diatomics being recently achieved \cite{Lee,OurRotations}.

Several proposals for  coherent control of vibration  in the deuterium molecular ion have been put forward, with similar underlying principles  \cite{Domhnallvibcontrol, NiikuraStoppingTheory,  Niederhausen}.
Firstly an ultrashort `pump' pulse can be used to induce strong field tunnel ionization of a D$_2$ target on a sub-vibrational timescale ($<$ 25 fs), creating a D$_2^+$ nuclear wavepacket in a coherent superposition of vibrational states.
Taking $t=0$ as the moment of creation of the molecular ion state, the subsequent vibration can be given by the eigenfunction expansion:
\begin{equation}
    \label{eqn:vibrational-expansion}
    \ket{\Psi(t)} = \sum_n a_n\ket{n}\exp(-\rmi E_n t/\hbar) \quad ,
    \end{equation}
where, $\{\ket{n }\}$ denotes the set of  vibrational eigenstates (discrete and continuous) of the molecular ion,
and  $\{E_n\}$ the corresponding energies. Each bound eigenstate component has an associated amplitude $\{a_n\}$ and evolves with frequency $\omega_n = E_n/\hbar$, where the population of each  $n$ state is $|a_n|^2$.

The evolution of the probability density is dictated by the beat frequencies ($\omega_n - \omega_{n'}$) between the eigenstates \cite{Feuerstein2003pra} with periods on the order of 20 - 30 fs for first order beats ($n-n' = \pm 1$).
The anharmonicity of the potential means that the
eigenstate components de-phase within a few vibration cycles \cite{Feuerstein2003pra}. The wavepacket will become spatially de-localised across the potential as time evolves but will re-form (to execute well defined oscillations) whenever the phases of the components in (\ref{eqn:vibrational-expansion}) match. That is, when the beats have re-phased. This quantum `revival' (at $t \sim 550$ fs) was recently measured
 \cite{ErglerSpatiotemp,Bryan} and well reproduced
by simulations \cite{Bryan,McKenna}.
In these experiments rotational effects in the molecular ion were found to be negligible \cite{Bryan} due to the large range of vibrational states occupied, each with different rotational constants \cite{Bocharova}.

Suppose that, after some time $\tau$ in the wavepacket evolution, a secondary (control) pulse is used to alter the vibrational wavepacket. It has been shown that an infrared control pulse can produce an AC Stark-effect that distorts the potential well (adiabatically) and hence modifies the wavepacket motion \cite{NiikuraStoppingTheory}. Depending on the wavepacket motion and position when the pulse is applied, a relative enhancement of lower vibrational populations may occur, i.e. vibrational cooling.
While  the  adiabatic picture provides a useful physical interpretation, it  does not fully explain the redistribution process.

\begin{figure}
\begin{center}
\includegraphics[width=\figwidth]{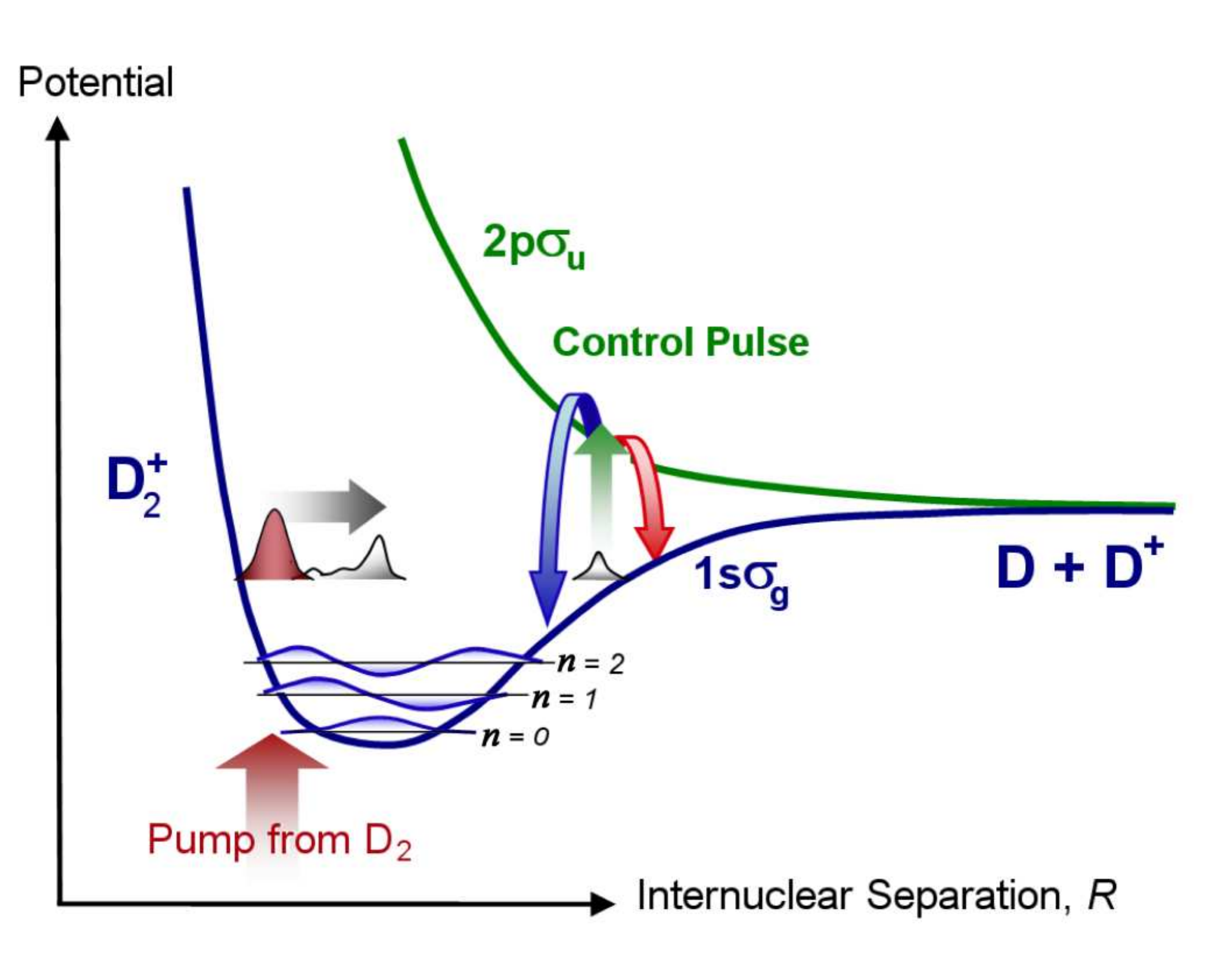}
\caption{\label{fig:schematic}
Schematic of  vibrational control of D$_2^+$ (not to
scale). An intense infrared pump laser interacts with  the D$_2$ target  in its ground vibrational state.  The ionization process creates  a  coherent superposition of D$_2^+$ 1s$\sigma_g$ vibrational modes displaced from the minimum of the well.  The wavepacket then  oscillates, due to the displacement, and disperses, due to  the anharmonicity of the D$_2^{+}$ state (silver arrow). The classical vibrational period (that is, the centre of mass of the oscillation)  is approximately 25  fs, while the vibrational modes will fully de-phase and re-phase (`revive')
after around 550 fs. The application of a secondary, control pulse initiates Raman transitions  via the
 2p$\sigma_u$ surface  creating Stokes (red arrow) and anti-Stokes (blue arrow) transitions.}
\end{center}
\end{figure}

To describe this, a Raman-type scheme has been proposed \cite{Domhnallvibcontrol,NiikuraStoppingTheory,Niederhausen}
where portions of the wavepacket may be transferred between the
1s$\sigma_g$ (bound) and 2p$\sigma_u$ (dissociative) potential, as seen in Figure \ref{fig:schematic}.  In recent
simulations, high-intensity pulses ($I\geq$ 10$^{14}$ W cm$^{-2}$) have been used to enforce vibrational squeezing or `quenching'
into a specific state \cite{Niederhausen} or to create a coherent 2 (or 3) state wavepacket \cite{Domhnallvibcontrol}. These studies were conducted with short delay times  ($\tau <100$ fs), where the first-order vibrational beat components are
de-phasing.

In this article, we show that intermediate intensity ($I \sim 5\times 10^{13}$ W cm$^{-2}$), few-cycle (5 fs) control pulses have important applications for coherent  manipulation. We find a
fascinating and hitherto unseen `chessboard' pattern in the redistribution. In a different outcome to previous
quenching/cooling studies, it is possible to selectively enhance
almost \textit{exclusively even} or \textit{exclusively odd}
numbered vibrational states. Thus the process relies on coherence in the energy domain, rather than wavepacket manipulation in the time-domain. 
It is found that the mechanism of this process relies on  constructive/destructive interference. This effect can be exploited at a fractional revival time to create strong-contrast
interference patterns.

\section{\label{sec:model} Simulations of Vibrational Control}

In this article, we simulate the control of a coherent D$_2^+$ vibrational wavepacket using a 5 fs, 5$\times 10^{13} $ W cm$^{-2}$  pulse. Referring to Figure 1, an ultrashort `pump pulse' (in  this case provided by a Ti:Sapphire  laser) interacts with  the initial D$_2$ target in its ground vibrational state.  The ionization process creates a  coherent superposition of D$_2^+$ 1s$\sigma_g$ vibrational states, as given in (\ref{eqn:vibrational-expansion}), displaced from the minimum of the potential well.  The wavepacket then  oscillates, due to the displacement, and disperses
due to  the anharmonicity of the D$_2^{+}$ state (silver arrow). The application of a secondary, control pulse initiates Raman transitions  via the 2p$\sigma_u$ surface (blue/red arrows), acting to redistribute the vibrational distribution of the wavepacket. This new vibrational distribution is studied as a function of the control pulse delay time, $\tau$.

The initial D$_2$ target can be assumed to reside in the ground vibrational state of the $X^1\Sigma_g$ potential. In the Franck-Condon (FC) approximation, the ionization event would simply  project this vibration function onto the manifold of vibrational states of the ion.
Regarding this pump process, it is well known (see \cite{Saenz2000,post01}, for
example) that the tunnelling-ionization rate  varies with bond-length, even for
the compact wavepacket presented by the D$_{2}$ ground vibrational state. This
can modify the simple FC approximation that assumes the wavepacket
is projected without modulation.
We have carried out simulations using modulated wavepackets (displaced/broadened/skewed) as well as the FC wavepacket and find that the `chessboard' interference effect we report here can be produced for each distribution.The FC  approximation will be used here to demonstrate the `chessboard' effect but it is not a necessary condition. A wavepacket with a range of populated vibrational states (such as those produced by intense infrared lasers) is sufficient. For initial distributions in a narrow range of vibrational states, the interference effects may not extend over as many states but will still occur. Indeed, for any experimental investigation the pump pulse parameters should be carefully chosen and compared to simulations for appropriate initial wavepacket conditions.

After the pump process, the FC wavepacket resides in the
ground electronic state of the D$_2^+$ ion. At intermediate intensities, it is sufficient to solve the time-dependent
equation for evolution on the  two lowest potential curves (1s$\sigma_g$ and 2p$\sigma_u$) within the Born-Oppenheimer approximation. With the infrared frequencies
considered here, these two low lying levels need only be considered as higher-lying electronic states are inaccessible.
 Thus the configuration space can be  partitioned into  $g$  and $u$ components. Then the  Hamiltonian, $H$,  can be partitioned into vibration, electronic motion, and laser interaction as follows:
 \begin{equation}
  H = T_R+H_e+V \quad .
  \end{equation}
  This has the matrix representation:
\begin{equation}
\fl
H =
\left(
\begin{array}{cc}
  T_R & 0 \\
  0   & T_R
\end{array}
\right)
+
\left(
\begin{array}{cc}
  \varepsilon_g(R) & 0 \\
  0   & \varepsilon_u(R)
\end{array}
\right)
+
\left(
\begin{array}{cc}
  0 & F(t) d(R) \\
  F(t) d(R)  & 0
\end{array}
\right) \quad  ,
\end{equation}
 where $T_R$ is the vibrational kinetic energy operator,
$\varepsilon_g(R)$ and $\varepsilon_u(R)$ are the potential energy
curves for the 1s$\sigma_g$ and 2p$\sigma_u$ electronic states
respectively, and $d(R)$ is the dipole moment between these
states \cite{Bates1951}. The external electric field $F(t)$ is
created by a Ti:Sapphire laser with $\lambda=790$ nm, modulated by a Gaussian
profile centred at $t=\tau$ (the delay time), with
full-width-half-maximum of $W$ (the duration), that is:

\begin{equation}
F(t) = F_0  \cos(\omega (t-\tau))  \exp\{- 2 \ln 2
(t-\tau)^2 /
W^2\}   \quad .
\end{equation}
The (cycle-average) intensity, $I$, of the pulse is related to the
electric
field amplitude, $F_0$, by the formula, $I = \frac{1}{2}\epsilon_0 c F_0^2$.
The evolution of the equations is computed by
a  symmetric, split-step algorithm \cite{Hermann1988}:
\begin{eqnarray}
\fl \ket{\psi(t + \Delta t)} =  \exp(- \rmi T_R \ \Delta t / 2) \exp(- \rmi H_e \ \Delta t / 2) \exp(- \rmi V(t) \ \Delta t )
\\ \nonumber
\times \exp(- \rmi H_e \Delta t / 2) \exp(- \rmi T_R \Delta t / 2)
\ket{\psi(t)} +  O(\Delta t^2) \quad  .
\end{eqnarray}
The error term in this expression arises partly from
the splitting (factorization) and from assuming
the Hamiltonian changes slowly over the time step (time-ordering error).
In practice, the time-step $\Delta t$ is chosen sufficiently small so that, $\Delta t(\partial H/\partial t) \ll H$.

The splitting and the use of a  uniform  grid means the highly-efficient  fast Fourier
transform \cite{FFTW05} can be employed. The electronic coupling term $V$ is diagonal in the radial
dimension, and a diagonalization of the $ 2 \times 2$ submatrix gives an efficient scheme
for propagation:
\begin{eqnarray}
\fl \exp
\left[ -\rmi \Delta t
\left(
\begin{array}{cc}
  0 & F(t) d(R) \\
  F(t) d(R)  & 0
\end{array}
\right) \right]
= \nonumber \\
\fl
{1 \over 2}\left(
\begin{array}{cc}
  \mathbbm{1}  &  \mathbbm{1}  \\
 - \mathbbm{1}  & \mathbbm{1}
\end{array}
\right)
\left(
\begin{array}{cc}
  \exp(-\rmi F(t) d(R) \Delta t) & 0\\
  0 & \exp(\rmi F(t) d(R) \Delta t)
\end{array}
\right)
\left(
\begin{array}{cc}
 \mathbbm{1}  & -\mathbbm{1}   \\
 \mathbbm{1}  &  \mathbbm{1}
\end{array}
\right) \quad  ,
\end{eqnarray}
where $\mathbbm{1}$ is the unit matrix. The vibrational populations
are  projections of the $g$-state wavepacket on the manifold,
$\{\ket{n} \}$,
while the
dissociation yield will be defined by the population in the $u$ state along with
the  $g$-state  continuum.

\section{\label{sec:explanation} The Chessboard}

\begin{figure}
\begin{center}
\includegraphics[width=\largefigwidth]{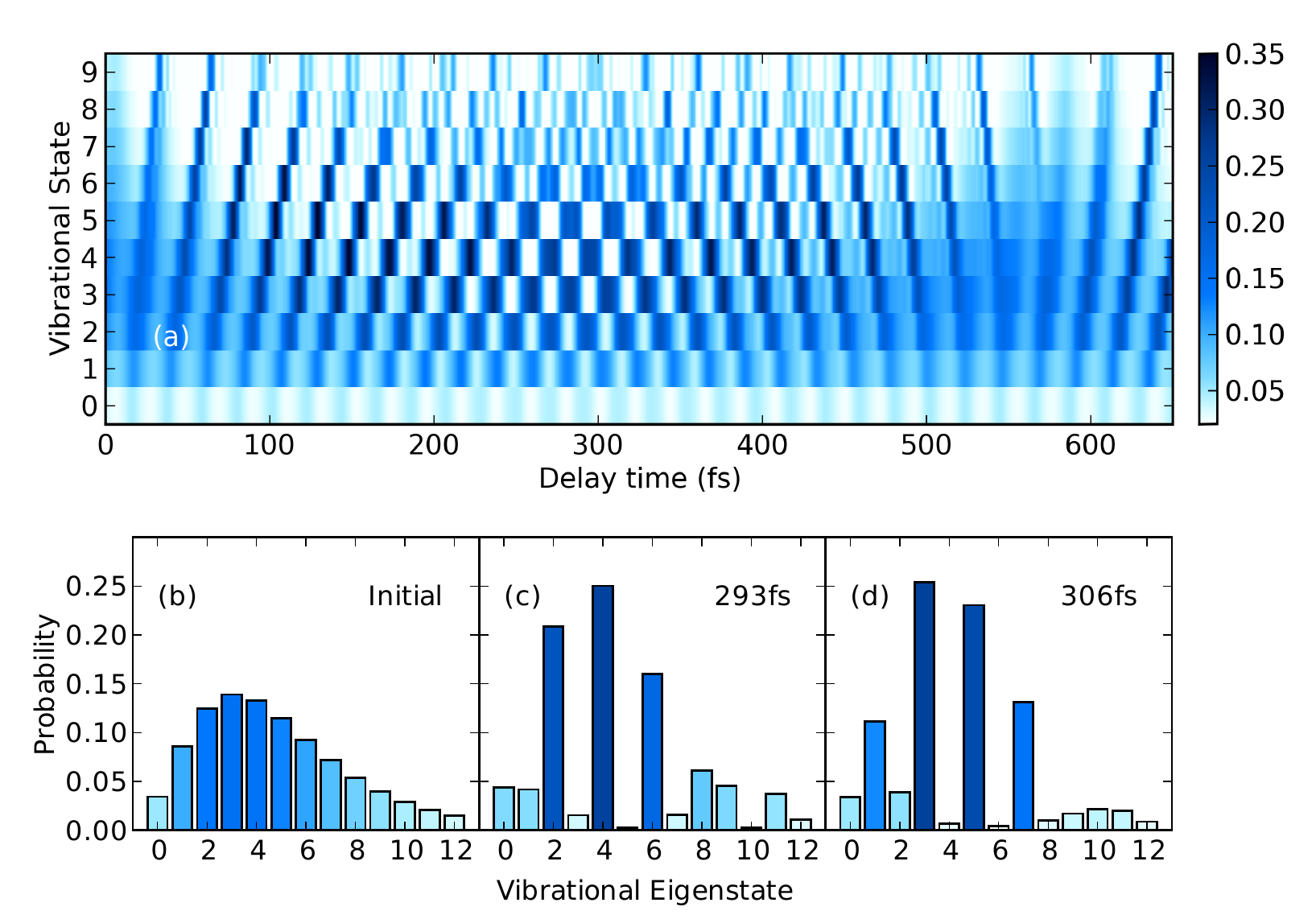}
\caption{\label{fig:bigfig} Vibrational distributions following a $\lambda=790$ nm, $W=5$ fs, $I=5\times 10^{13}$ W cm$^{-2}$ control pulse, as a function of delay time $\tau$.
(a)  The colour  density represents the relative population of each level. (b) Initial probability distribution. For short delay times, $\tau<$ 100 fs, and near the revival, $\tau \sim 550$ fs, the control pulse redistributes broadly with only slight enhancement in any single level. In contrast, the `chessboard' pattern (centred around the fractional-revival $\tau \sim 300$ fs) shows an interference pattern in which even or odd numbered states are annihilated. For $\tau$ = 293 fs and $\tau=306$ fs, cuts through the colourmap are shown in (c) and (d).}
\end{center}
\end{figure}

The vibrational populations,  following the application of a
short control pulse ($W= 5$ fs), with
$I= 5\times 10^{13}$ W cm$^{-2}$, were calculated for a range
of delay times $ 0$ fs $\leq \tau \leq 650$ fs. The results are presented  in Figure \ref{fig:bigfig}. The colour density in Figure \ref{fig:bigfig} (a) represents the final population of a
vibrational level with respect to the control pulse delay.
The distributions  resulting from $\tau$ = 293 fs and $\tau=306$ fs are extracted from Figure \ref{fig:bigfig} (a) and
displayed as the  bar charts \ref{fig:bigfig} (c) and \ref{fig:bigfig} (d), respectively.  The initial  probability
distribution is shown  in Figure \ref{fig:bigfig} (b).

In previous studies, the  modulations in population
had been  attributed to the classical
vibrational period of each level \cite{Niederhausen}. Here a description for the mechanism will be given in terms of interference processes  in the energy 
domain regulated by the beat frequencies.

For short delays ($\tau < 50$ fs) the anharmonic correction
is small, and dephasing is not significant.
Under such conditions, the  control pulse spreads the
population evenly  with only  slight enhancements/deficits.
In the region $\tau \sim  100$ fs, enhancement is concentrated in a specific level (e.g. $n$ = 4 at $\tau = 100$ fs). These effects are repeated around the wavepacket revival ($\sim$ 550 fs), were the wavepacket begins to re-form and then de-phase again.
Higher-intensity control pulses can be used to optimize
state-selective control \cite{Domhnallvibcontrol,Niederhausen} and drive population towards one or more specific levels.

The most striking feature of  Figure \ref{fig:bigfig} (a) is the
alternating light and dark squares, suggestive of a  `chessboard' pattern, centred around $\tau \sim 300$ fs. At this fractional revival time,  the even-numbered 
vibrational states will be in phase with
each other but in anti-phase with the 'odd' states.
This  phase relationship can be exploited to produce
interference effects. The effect is illustrated in Figure \ref{fig:bigfig} (c), where at $\tau =$ 293 fs the non-neglible populations are \textit{exclusively} in \textit{even} numbered states. Likewise at $\tau =
306$ fs only \textit{odd} numbered states are populated, with
$n= 1,3,5,7$, strongly favoured.

\begin{figure}
\begin{center}
\includegraphics[width=\figwidth]{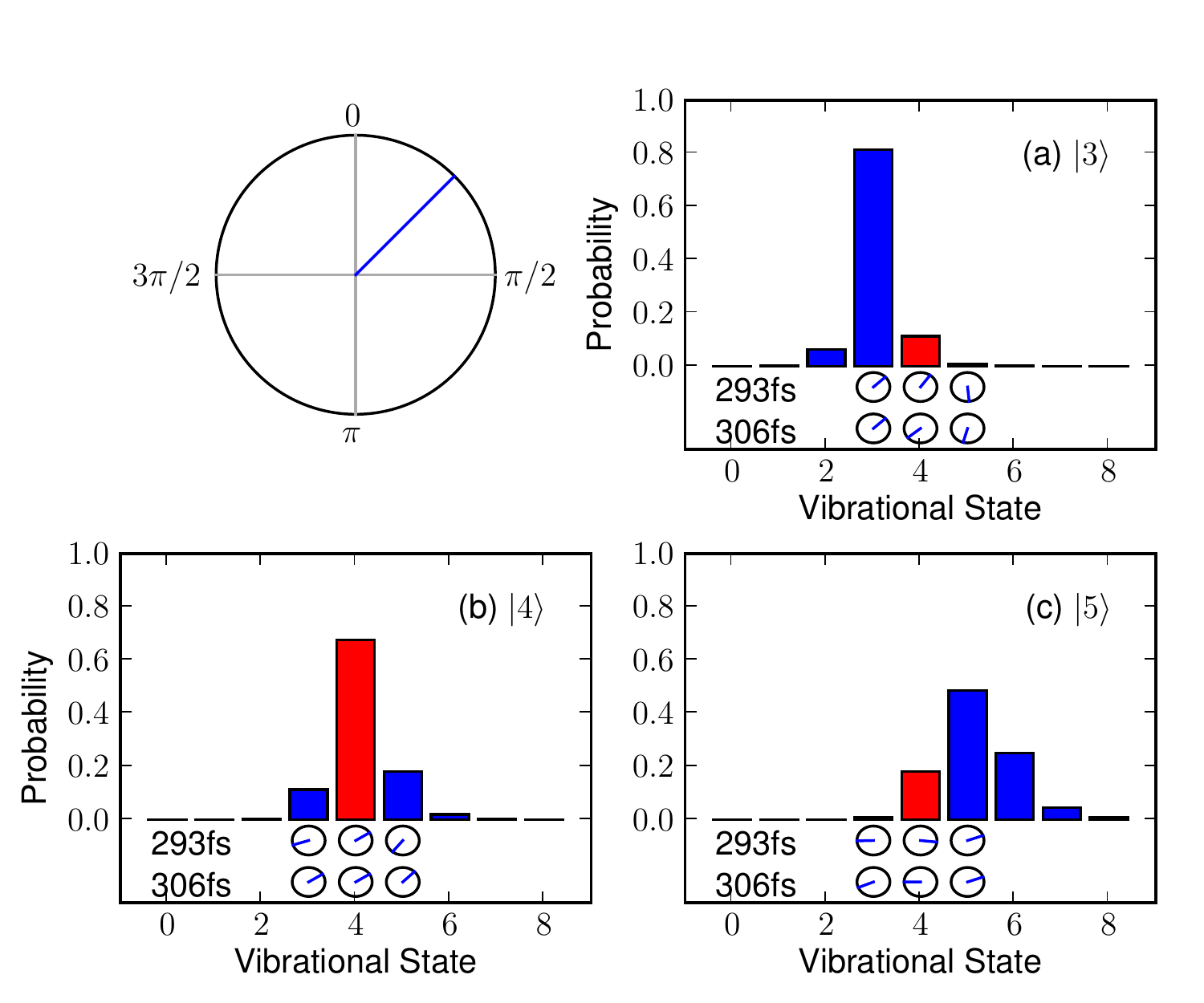}
\caption{\label{fig:clockfig}
Vibrational transfer starting from  (a) $n=3$, (b) $n=4$ and (c) $n=5$ by
a 5 fs, $\lambda=790$ nm, $I=5\times 10^{13}$ W cm$^{-2}$ control
pulse.  The bar charts represent the probabilities $\vert
a_n\vert^2$ while
the clocks display the  phase of  the coefficients $a_{n}$
after the pulse for $\tau = 293$ fs and
$\tau = 306$ fs. The distributions  are concentrated  around the
initial state, while the initial state is phase-advanced. This
illustrates that a coherent sum of the final distributions will
 depend strongly on $\tau$. For the reader's convenience, the  contributions into $\ket{4}$ are highlighted in red, see text for details.}
\end{center}
\end{figure}

The effect can be understood by the destructive/constructive
interference between
nearest-neighbour states.
Suppose we consider the effect of the pulse on isolated  states $n=3,4,5$, separately.
In Figure \ref{fig:clockfig} the final amplitudes and phases for
each state
are illustrated by bar charts and clocks respectively.
The clocks display the phase shifts for the wavepacket population
moved into each level, with the relative phase shift defined with
respect the expected phase for an unperturbed state.
All clocks will remain at 12 o'clock
in the absence of any perturbation.
In each case, starting with a single vibrational state,  the final
probability distribution is independent
of $\tau$. We also note a selection rule favouring neighbouring
levels. To illustrate this, let us focus on the state $n=4$. For
the reader's convenience, the bar charts showing the contributions
into $\ket{4}$ are highlighted in red in Figure \ref{fig:clockfig}.
  At the delay time $\tau = 293$ fs, the contributions
to $\ket{4}$ from all three initial states have roughly the same phase, i.e. they interfere
constructively. In contrast, at $\tau = 306$ fs, the
contributions from $\ket{3}$ and $\ket{5}$
are almost opposite to the contribution from $\ket{4}$, giving a destructive interference. For the
final population in $\ket{5}$, the situation is opposite, yielding destructive and constructive
interference for $\tau = 293$ fs and 306 fs respectively. This pattern is repeated
so that even numbered states get enhanced and odd numbered states get quenched for $\tau = 293$ fs,
and vice-versa for $\tau = 306$ fs giving rise to the
chessboard effect seen in Figure \ref{fig:bigfig}.

Some  previous work on  wavepacket engineering focussed on 
manipulation in configuration space. 
The  chessboard is an effect that  pertains to 
engineering in the energy dimension. We note that, 
although  the chessboard state is highly regular in energy space, the wavepacket in configuration space is
still quite irregular as time progresses.

\section{Model}

A simple mathematical model of the interference can be obtained from perturbation
theory. As a preliminary comment, the hierarchy of timescales in this system are worth
noting, from the fast electronic motion to the medium-speed optical field,
through to the comparatively `slow' vibrational motion.
The `electronic time' is of the order of the dissociation energy of
the molecule, with an associated angular
frequency: $\sim 25 \times 10^{15}$ rad s$^{-1}$
$\approx 0.6 $ a.u., while the optical cycle time
is $\omega_{\rm opt}  \sim 2.4 \times 10^{15}$ rad
s$^{-1}$ $\approx 0.06 $ a.u.,
and finally the vibration time $\omega_e \sim 0.3 \times 10^{15}$
rad s$^{-1}$ $\approx 0.007$  a.u.
Consequently the pulse interaction is sudden compared to the
vibrational timescale, but adiabatic in the context of the
electronic motion.

The  effect of a laser pulse after a delay $\tau$ can now be
considered. A simple mathematical model can be constructed
where  the the modulated sinusoidal signal
of the laser pulse is considered as a
sequence of alternating square waveforms. The
expansion
\begin{equation}
\label{eqn:full-expansion}
\ket{\psi(t)} = \sum_{{\gamma,n}} a_{\gamma,n} (t) e^{-\rmi E_{\gamma,n}t} \ket{\gamma, n}   \qquad ,
\end{equation}
can be proposed as the  solution to
\begin{equation}
\label{eqn:tdse}
 \rmi  {\partial \over \partial t} \ket{\psi(t)}= H(t) \ket{\psi(t)}  \qquad ,
\end{equation}
where $\gamma \in \{ g, u \}$  is the electronic state with corresponding
vibration state  (bound and/or dissociative) $n$.
 Equation (\ref{eqn:tdse}) can be written as a set of integral
equations ($t \geq t_{0}$):
\begin{equation}
a_{\gamma,n} (t)
 = a_{\gamma,n} (t_{0})- \rmi \sum_{\gamma',n'}^{}  \int_{t_{0}}^{t}
dt'  a_{\gamma',n'} (t') V_{\gamma,n;\gamma', n'}(t')
  \exp \left[  \rmi \Delta_{\gamma,n;\gamma', n'}t' \right]  \quad ,
\end{equation}
where $\Delta_{\gamma,n;\gamma', n'}= E_{\gamma,n} - E_{\gamma', n'} $ and
the coupling potential $V$ is expanded in the basis set in
equation (\ref{eqn:full-expansion}). In the two-state approximation 
the only non-zero element  has the form
\begin{equation}
  V_{g,n;u, k}(t')  =   V_{u,k;g, n}(t')  = 
  F(t)  \langle k \vert d(R) \vert n \rangle  \qquad ,
\end{equation}
where $ \vert k \rangle $ is an energy-normalised continuum function of the
$u$-state, with wavenumber $k$, and $\vert n \rangle$, is a vibrational eigenstate 
(bound or continuum) of the $g$-well. 
 
The photon energy ($\sim$ 0.06 a.u. in this case)
is below the single-photon resonant dissociation for all states of the $g$ potential that are shown in Figure \ref{fig:bigfig}, and  for laser intensities lower than $10^{14}$ W cm$^{-2}$, the  upper electronic level $u$ acts as a virtual
state. 
The expansion coefficients of  (\ref{eqn:full-expansion}) of the $g$-state
can then be calculated to second-order accuracy.

\begin{eqnarray}
\fl a_{g,n} (\tau+W)
     & =   &  a_{g,n} (0)+(-\rmi)^2 \sum_{k,n'}
  \int_{\tau}^{\tau+W} dt' \exp \left[  \rmi
\Delta_{g,n;u, k}t' \right]
  V_{g,n;u,k}(t')  \\ \nonumber
  & \qquad \times &
    \int_{\tau}^{t'} dt'' \exp \left[  \rmi \Delta_{u,k;g, n'}t'' \right]
  V_{u,k;g, n'}(t')
  a_{g,n'} (0)   \quad .
\end{eqnarray}

For simplicity, consider the pulse to be applied impulsively
 after a delay $\tau$, for a duration $W'$ (corresponding to half an optical cycle),
 with amplitude $F_0$.
A further simplification can now be made in the summation over
the intermediate states. The  'closure approximation' replaces
the spectrum of intermediate
energies by a  constant,  $E_{u,k}=\bar{E}$,
representative of the dominant (resonant) channels.
 In fact the choice of $\bar{E}$ 
can be used as a free parameter if desired, but it 
must lie in the continuum.  In  the spirit of the
 approximation, we choose $\bar{E}$ as the
threshold of dissociation in the $u$ state. also reflecting 
the enhanced density of states at threshold.  Then
taking this as the reference point for the zero of energy, we simply
set  $\bar{E} = 0$. The closure
(completeness) relation can then be applied so that:
\begin{equation}
\sum_{k}  \langle n \vert d (R) \vert k \rangle \langle k \vert d(R) \vert n' \rangle 
=  \langle n \vert d^2(R) \vert n' \rangle \equiv d^2_{n,n'} \quad .
\end{equation}
The index $g$ may now be dropped for convenience and we have:
\begin{equation}
\fl a_{n} (\tau+W')
      = a_{n} (0)-\sum_{n'}
     a_{n'}(0)  d^2_{n,n'}F_0^2
  \int_{\tau}^{\tau+W'} dt' e^{ \rmi E_n t' }    \int_{\tau}^{t'}
dt''  e^{- \rmi E_{n'}t''} \quad ,
 \end{equation}
and this may be evaluated without further approximation.

\begin{figure}
\begin{center}
\includegraphics[width=\figwidth]{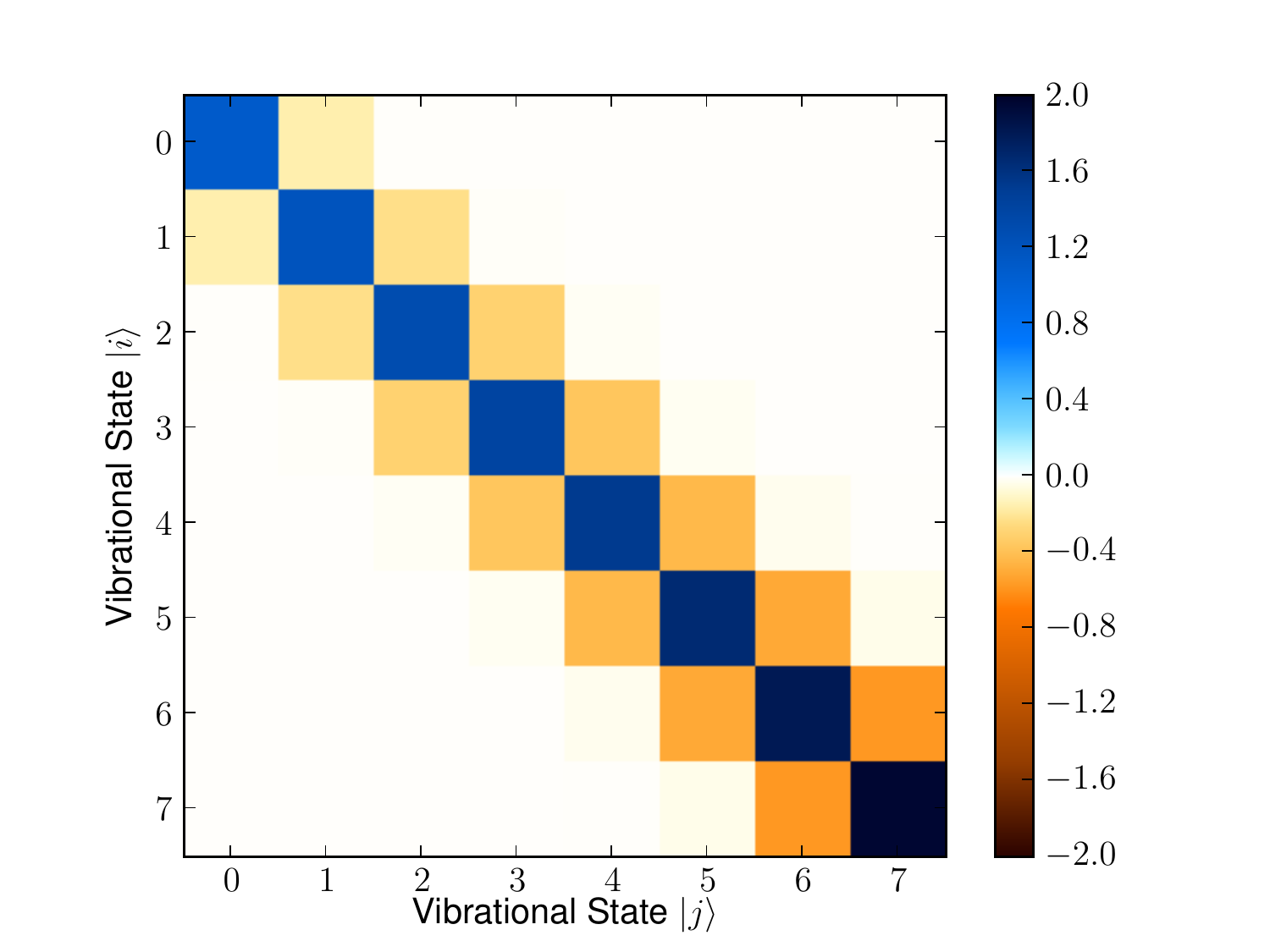}
\caption{\label{fig:couplingfig}
Second-order coupling strength between vibrational eigenstates
during the laser pulse $(d^2)_{ij}$. Notice the rapid decrease in
coupling
strength as $|i-j|$ grows leading to an approximate selecton rule
$i-j \in \{-1, 0, 1\}$ for intermediate fields This  arises
from the linear form of the function $d(R)$. Also notice that the
coupling strength increases  with vibrational levels, giving a rise
to a slight heating, i.e.propensity towards  higher vibrational
levels.
}
\end{center}
\end{figure}

As seen in Figure \ref{fig:couplingfig}, the matrix element $d^2_{n,n''}$ provides an approximate selection rule:
$n'=n,n\pm 1$, so that the expression can be reduced even further. Moreover,
since  $E_n W' \gg 1$, then
 the solution can be expressed in the simple form:
\begin{eqnarray}
 \label{eqn:final-population}
 \fl a_{n} (\tau+W') \approx &
  a_{n} (0)   \left(  1- \rmi \kappa_{nn} \right)
  -i \kappa_{n,n-1}   a_{n-1} (0)  e^{\rmi(E_n-E_{n-1})\tau}
  \\
 & -i \kappa_{n,n+1}   a_{n+1} (0)  e^{\rmi(E_n-E_{n+1})\tau} 
 \quad , \nonumber
\end{eqnarray}
where,
\begin{equation}
 \kappa_{n,n'} = \left( { F_0^2  d^2_{n,n'} \over E_{n'} }  \right)
 \left(
{ \exp [ i(E_n-E_{n'})W']-1 \over  i(E_n-E_{n'})}
 \right)  \quad .
\end{equation}
In the case of a very short impulse, corresponding to a half-cycle,
$(E_{n}-E_{n-1} )W' \ll 1$, and:
\begin{equation}
 \kappa_{n,n'} \approx \left( { F_0^2 W' \over E_{n'} }  \right) d^2_{n,n'} \quad .
\end{equation}
The term $\kappa_{n,n}$ simply reflects the change in phase due to the
quadratic  Stark effect. Since  this is a
second-order effect, the  sign of $F_0$ has no consequence, and
$\kappa_{n.n}  < 0$, and hence there is a phase advance. This is illustrated
in Figure \ref{fig:clockfig} where the parent state is phase shifted forward, with the
excited state $n=5$ more advanced than the lower-energy state $n=3$, in agreement with this expression for the energy dependence.
We note that, in this model, $\kappa_{n,n'}$
is not sensitive to the sign of $F_0$, that is, to the carrier envelope phase. Thus
a sequence of alternating half-cycles creates the same effect as a full-wave rectified pulse.

Consider the effect of one half-cycle.
The equations above have the simple explanation that
$a_{n} (\tau+W')$ is created by an interference pattern with its immediate neighbours.
This pattern can be regular if the terms on the right hand side are
coherent. In order to create strong destructive interference, the
terms  ($\kappa_{n,n-1} a_{n-1}(0) + \kappa_{n,n+1} a_{n+1}(0)$) must
be approximately
the same size as the remaining term $a_n(0) (1 - \kappa_{n,n})$ for $n \in 2 \ldots 6$. This gives an optimal value of $F_0$ and $W'$. Next, the phases of the
two interference terms must be equal, which implies
the condition (for all $n$):
\begin{equation}
 \label{eqn:phase-half-revival}
(E_n-E_{n-1}) \tau = -(E_{n+1}-E_{n}) \tau   \quad , \quad  {\rm
mod}(2\pi) \quad .
\end{equation}
A rough estimate of this condition on $\tau$, can be made by  the anharmonic expansion, $
E_n \approx -D_e+\hbar \omega_e(n+\textstyle{1 \over 2} ) -\hbar \omega_ex_e(n+\textstyle{1 \over 2} )^2 $,
where $D_e$ is the dissociation energy.
This is satisfied during the fractional revival around $\tau \approx \pi / (2\omega_e x_e)  \sim 280$
fs. Finally, these terms  must be out of phase with $a_n$ to
create destructive interference. The Stark effect  (\ref{eqn:final-population}) creates a
phase shift,
$-\kappa_{n,n}$, for the state $n$. Thus  in order to get destructive interference, the requirement is then
\begin{equation}
 \label{eqn:phase-shift}
 \kappa_{n,n} = (E_n - E_{n-1}) \tau + \pi/2  \quad , \quad  {\rm
mod}(2\pi) \quad .
\end{equation}
If destructive interference occurs for even $n$,
then according to equation (\ref{eqn:phase-half-revival}), constructive
interference will be observed for odd $n$.
Furthermore, if destructive interference is observed for a state at $\tau = \tau_0$, constructive interference
will be observed for $\tau = \tau_0 + \pi / \omega_{e}
\approx \tau_0 + 11$ fs.
This is a qualitative explanation, and is limited by the crudeness
of the closure approximation.  Further, the assumptions we make for
the pulse are simple. Nevertheless, it appears that this simple
model  explains the main features that we observe in the numerical
simulations.

\section{Proposed Experimental Technique for Observing the Chessboard}

\begin{figure}
\begin{center}
\includegraphics[width=\figwidth]{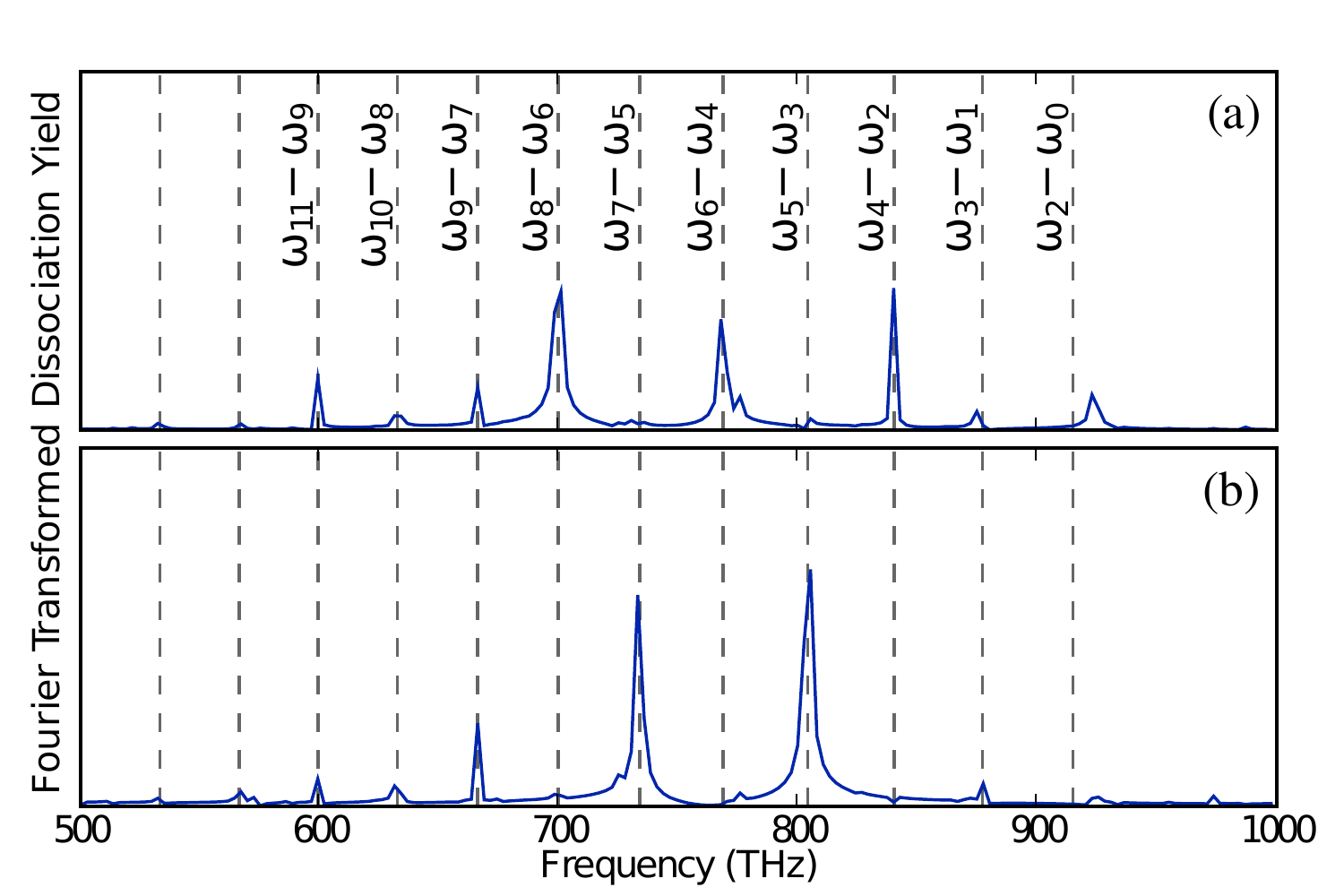}
\caption{\label{fig:FFTfig_expt_observable} Spectral density for the dissociation yield correlation function. The coherent
D$_2^+$ wavepacket is controlled by a 5 fs, $\lambda=790$ nm,
 $I=5 \times 10^{13}$ W cm$^{-2}$ pulse at $\tau = 293$ fs and $\tau=306$ fs for plots (a) and (b) respectively. The probe pulse is, $\lambda=790$ nm,  $I=4 \times 10^{14}$ W cm$^{-2}$, duration 5 fs.  Fourier analysis of  the dissociation
signal returns the vibrational beats $\omega_{n+2}- \omega_{n}$ present in the wavepacket motion in each case.
 The probe delay time ranges from  310 fs to 4000 fs.
A shorter range of times, 310 fs to 1000 fs for example,  gives similar features. In this case  the peaks are broadened but still  well resolved.}
\end{center}
\end{figure}

Having simulated the chessboard pattern and explained the
underlying mechanism, attention can be turned to techniques for
experimental observation of this fascinating effect.
In recent experiments \cite{ErglerSpatiotemp,Bryan}, coherent vibrational wavepacket motion in D$_2^+$ has been initiated and imaged in a pump-probe configuration using short intense-field pulses. In these studies the vibrational
wavepacket motion, and notably the vibrational revival, has been imaged by using a probe pulse to enforce photodissociation or Coulomb explosion of the vibrating molecule across a range of probe delay times.

Fourier analysis of the fragmentation yield can return a measure of both the temporal and spatial nature of the wavepacket motion, which can ultimately lead to full characterization of the molecular motion \cite{Feuerstein2007prl}. In particular, the
beat frequencies that  dictate the motion of the
probability density can be extracted from the
photodissociation signal.

Thus it may be possible to conduct a pump-control-probe experiment where a
new coherent vibrational distribution (created at a specific control pulse delay
time) may be monitored by recording the total dissociation yield as a function
of probe delay time ($\tau'$) with subsequent spectral analysis of the signal.

It is important that for modelling any such experiment, the pump process is well
characterised. For pulse intensities of $\sim$ 1$\times 10^{14}$ W cm$^{-2}$ the
$R$-dependence (see \cite{SAENZ} and references therein) of the pump process has
been considered in recent studies of D$_2^+$ wavepackets
\cite{Niederhausen,Feuerstein2007prl,Thumm2008}, and indeed in some recent
experiments \cite{ErglerSpatiotemp} the results appear to deviate from an FC
wavepacket.
It is noted here that the R-dependence of the pump process is sensitive to pulse
intensity \cite{Urbain} and, as pulse intensity increases, the resulting ion may
tend towards an FC distribution \cite{Urbain},
However, if the intensity is set too high there is a probability that direct
fragmentation of the D$_2$ target may occur in the initial pump pulse, clouding
the detection of any bound wavepacket dynamics from the control or probe pulse
interactions.

At this point, it is instructive to observe that in other recent  experiments
\cite{Bryan,McKenna} with pump pulses ($\lambda=790$ nm) of duration $10-15$ fs,
and intensities of $5-8 \times 10^{14}$ W cm$^{-2}$ the wavepacket dynamics are
in fact well reproduced by Franck-Condon simulations. In these experiments the
problem of direct D$_2$ fragmentation from an intense pump pulse is overcome by
orienting the pump pulse polarisation perpendicular to the detection axis,
ensuring that any direct fragmentation arising from this pulse goes undetected.

In this context, the proposed experimental technique for identifying the
chessboard effect has been simulated for an FC wavepacket. This was done by
fixing the control pulse at a desired $\tau$ value, and introducing an intense
probe pulse (5 fs duration, $I=4\times 10^{14}$ W cm$^{-2}$) at  a variable
delay time $\tau'$. The dissociation yield was deduced by subtracting the
remaining bound wavepacket population from the initial norm of the wavepacket.
The Fourier transform (energy spectral density)  of the dissociation signal is
shown in Figure \ref{fig:FFTfig_expt_observable} for control delay times of (a)
$\tau=293$ fs and (b) $\tau=306$ fs. The calculations represent the signal
correlation over very long times (4000 fs). We have performed the same analysis
over a much shorter time (1000 fs) and, although the features are broadened, the
signatures are still clearly visible.

The individual states in the vibrational distribution are not resolved in
isolation but rather the peaks in Figure \ref{fig:FFTfig_expt_observable} (a)
correspond to the beat frequencies between even states. These are notably absent
in Figure \ref{fig:FFTfig_expt_observable} (b), where only the odd numbered
eigenstate beat frequencies are  observed, in keeping with the expected
vibrational distributions (from Figure \ref{fig:bigfig}). The amplitude of the
beat $\omega_{n+2} -\omega_{n}$ in the wavepacket motion is given by
$|a_na_{n+2}|$ . The Fourier transform technique will not extract
the absolute population values,  $|a_na_{n+2}|$ but rather, will provide insight
into which states are significantly populated. This is a useful technique for
experimentally identifying the preferential creation of either \textit{even} or
\textit{odd} numbered states. It should be noted however that periods of the
$\omega_{n+2} -\omega_{n}$ beats are typically  $10 - 12$ fs and thus for any
experimental verification, it is imperative that pulses as short as 5 fs are
used.

In principle, this technique using the photodissociation signal could be a
useful method for identifying the main components of the wavepacket, and with
state-of-the-art pulses now available for durations less than 10 fs, this should
be achievable in the laboratory in the not too distant future. Indeed, with
continuing advances in intense-field laser-pulse experiments, such an even (or
odd) state wavepacket (evolving with $\sim$ 12 fs oscillations) may even be
 characterized by using a high-intensity probe pulse and a very recently
reported `time-series analysis' of the Coulomb explosion signal
\cite{Thumm2008}, where full reconstruction of the wavepacket has been proposed.

\section{Conclusion}

In this article, the application of an intermediate intensity ($5\times 10^{13}$
W cm$^{-2}$) few-cycle (5 fs) pulse to control a coherent vibrational wavepacket
in D$_2^+$ has been investigated. The redistribution of vibrational levels
proceeds by a Raman transfer between near-neighbour levels. This creates
interferences such that it is possible to produce strong-contrast fringes
of the even/odd numbered vibrational levels. The effect is a property of the
molecular ion and does not require a highly-selective  ionization process. All
that is required is that a band of vibration states is populated. Simulations
using the Franck-Condon approximation are shown for illustration but this is not
a necessary condition.

The `chessboard' pattern that emerges  around $\tau \sim 275 - 300$ fs provides
a pathway for creating almost \textit{exclusively even} or \textit{exclusively
odd} numbered vibrational distributions in a coherent wavepacket. Thus the
process relies on coherence in the energy domain, rather than wavepacket
manipulation in the time-domain. The subsequent propagation of such a created
wavepacket may be probed via photodissociation and here it has been shown that
spectral analysis of the dissociation data may serve to identify the populated
states. This provides a pathway for using currently available ultrashort laser
pulse technology to investigate this `chessboard' effect.

\ack

C R Calvert  and R B King wish to acknowledge funding from Department of
Employment and Learning (NI).

\end{document}